\documentclass[preprint,prb,amsmath]{revtex4-2}

\usepackage[pdftex]{graphicx} 
\usepackage{color}
\usepackage{bm}
\usepackage{epsfig}
\usepackage{latexsym}
\usepackage{amsmath}
\usepackage{nameref,hyperref}
\usepackage{cases,setspace,amsmath}
\usepackage{wasysym}

\makeatletter
\let\old@makecaption=\@makecaption
\usepackage{subcaption}
\let\@makecaption=\old@makecaption
\makeatother
\usepackage{blindtext}

\usepackage{hyperref}

\def\av#1{{\langle  #1 \rangle}}
\def\be{\begin{equation}}
\def\ee{\end{equation}}
\def\bea{\begin{eqnarray}}
\def\eea{\end{eqnarray}}
\def\bsn{\begin{subnumcases}}
\def\esn{\end{subnumcases}}

\begin{document}

\title{Scaling regimes in slow quenches within a gapped phase}
\author{Lakshita Jindal}
\email{lakshita@jncasr.ac.in}
\author{Kavita Jain} 
\email{jain@jncasr.ac.in}
\affiliation{Theoretical Sciences Unit, \\Jawaharlal Nehru Centre for Advanced Scientific Research, \\Bangalore 560064, India}

\date{\today}
\begin{abstract}
  We consider the finite-time quench dynamics in the quantum transverse field Ising model which exhibits a second order phase transition from a paramagnetic to a ferromagnetic phase, as the transverse magnetic field is decreased. These dynamics have been thoroughly investigated in previous studies when the critical point is crossed during the quench; here, we quench the system from deep in the paramagnetic phase to just above the critical field so that the system remains in the gapped phase throughout the quench duration. 
  On linearly quenching the infinitely large system, we find that the behavior of mean longitudinal defect density and mean transverse magnetization at the end of the quench falls into three distinct scaling regimes as the quench time is increased.  For sufficiently small quench times, these observables remain roughly constant, but for larger quench times, a crossover occurs from the Kibble-Zurek scaling law to the quadratic quench rate law when the Kibble-Zurek time is of the order of relaxation time at the final quench field. 
These results are shown analytically using power series and uniform asymptotic expansions of the exact solution of the model, and also compared  with an adiabatic perturbation theory in the third regime. We find that the above mentioned scaling regimes hold for quenches within the ferromagnetic phase also, and provide a general scaling argument for crossover from the Kibble-Zurek regime to an adiabatic regime for slow quenches within a gapped phase.
  \end{abstract}
  
\maketitle

\clearpage
\section{Introduction}


Quenches across the critical point have been studied extensively for several decades in condensed matter systems exhibiting quantum phase transition to understand their nonequilibrium behaviour \cite{sachdev1999quantum}. If the quench is instantaneous or sufficiently rapid, the system is not in equilibrium state at the end of the quench, and one is interested in understanding if and how the system relaxes towards its ground state \cite{essler2016quench}. Quenches occurring at a finite rate (dubbed as slow quenches) are of great interest as they allow one to  interpolate between the two extreme limits of sudden quenches and adiabatically slow change of parameters, and open various possibilities of new features in the nonequilibrium dynamics \cite{Dziarmaga01112010,polkovnikov2011colloquium,chandran2012kibble,dutta2015quantum}. 


From the theory of critical phenomena, it is known that quantum (and classical) systems exhibiting a second order phase transition between a disordered and ordered phase are characterized by diverging length and time scales near the critical point. Then, if the system is quenched slowly in the vicinity of the critical point, due to the critical slowing down, the system can not keep up with the changing control parameter and falls out of equilibrium which leads to the absence of adiabaticity, and more defects than in the ground state. The excess defect density decays as a power law with the quench time and the scaling exponent can be obtained by an argument which was first proposed by Kibble to describe the symmetry breaking of early universe \cite{kibble1976topology, kibble1980some} and later extended by Zurek to condensed matter systems \cite{ zurek1985cosmological,zurek1996cosmological}.

Since then, the slow quench problem has been studied theoretically in a variety of classical \cite{brey1994dynamical, biroli2010kibble, krapivsky2010slow, jain2016critical, Priyanka2021, mayo2021distribution, kim2022nonequilibrium, jindal2024kibble} and quantum \cite{zurek2005dynamics,polkovnikov2005universal,dziarmaga2005dynamics,   kolodrubetz2012nonequilibrium,bialonczyk2018one,schoenauer2019finite,puebla2020kibble, bialonczyk2020locating, schmitt2022quantum,roychowdhury2021dynamics,soriani2022three, kou2023varying} models and for various quench protocols \cite{das2006infinite, sen2008defect, puskarov2016time, suzuki2023kibble}, and   
 in experiments \cite{del2014universality, keesling2019quantum, osti2340752} on a wide variety of systems such as Bose gases \cite{beugnon2017exploring, navon2015critical, goo2022universal}, ion crystals \cite{de2010spontaneous}, quantum computers \cite{gardas2018defects, king2022coherent}, semiconductors \cite{wang2014quantum}, etc. to test the predictions of Kibble-Zurek (KZ) theory, and found to be in good agreement. However, recent work has explored the limitations of KZ theory, and shown that if quench is sufficiently fast  \cite{chesler2015defect, zeng2023universal} or noise field is applied to the system \cite{dutta2016anti, iwamura2024analytical, cui2020experimentally}, the KZ argument does not apply and new scaling laws may hold. 

Besides the KZ theory, for large quench times where the system is expected to be close to the equilibrium state, adiabatic perturbation theories (APTs) \cite{polkovnikov2005universal,rigolin2008beyond,de2010adiabatic,de2013microscopic,rigolin2014degenerate} have been developed to understand the  approach to the ground state. 
For quenches within the gapped phase, the excess defect density has been shown to vanish in a universal fashion, as the square of quench rate 
and, interestingly, the KZ scaling for quenches from disordered to the ordered phase has also been obtained \cite{polkovnikov2005universal,de2010adiabatic}. 

Here, we consider a paradigmatic model of quantum phase transitions, namely, the transverse field Ising model (TFIM) \cite{ pfeuty1970one, pfeuty1971ising} (for a recent pedagogical review, see \cite{mbeng2024quantum}), and focus on finite-time quenches when the system initially prepared in the ground state at large transverse field is quenched just above the critical point so that the system remains in the paramagnetic phase throughout the quench duration.     
In previous work, KZ scalings have been observed when the system prepared deep in the disordered phase is quenched  either to the critical point \cite{bialonczyk2018one} or the ordered phase \cite{zurek2005dynamics,polkovnikov2005universal,dziarmaga2005dynamics}; in our study, we find that KZ scalings  hold even for quenches within the gapped phase, which, to our knowledge, has not been noted before. We find that when the KZ time is small compared to the equilibrium relaxation time at the final quench field, the excess defect density decays according to the KZ scaling law, and on increasing the quench time, a crossover occurs to the quadratic scaling in quench rate.

In the following sections, we obtain analytical expressions for excess mean longitudinal defect density as well as excess mean transverse magnetization in these regimes using the relevant expansion of the exact solution of the TFIM. For large quench times (where quadratic quench rate law holds), we also consider an adiabatic perturbation theory \cite{rigolin2008beyond} which yields only analytic dependence on quench time, and has been used to investigate quenches within the ferromagnetic phase \cite{maraga2014nonadiabatic}; we find that the uniform asymptotic expansion of the exact solution and the APT match well, and interestingly, an explicit expression for the amplitude of the quadratic decay can be found using the former expansion. Before proceeding to the specific model, in the following section, we first give a general argument to understand the crossover in the scaling for slow quenches within the gapped phase.

\section{Scaling arguments}
\label{scar}

Consider an infinitely large quantum system that exhibits a second order phase transition at the critical point $g_c$  between an ordered phase ($g < g_c$) and a disordered phase ($g > g_c$) when the control parameter $g$ is varied.
If the system initially equilibrated to $g_i > g_c$ is quenched in  time $\tau$ to $g_f < g_i$, 
the system equilibrates  to the ground state at $g_f$ if the quench time is infinitely long otherwise an observable such as mean density of defects at the end of quench is more than that expected in the ground state. For $g_f \leq g_c$ (that is, if the system is quenched to or through the critical point), with increasing quench time, the  
residual density of defects, $\delta {\cal D}$ (typically) decays as a power law, and the decay exponent can be obtained using the KZ argument as follows \cite{kibble1980some,zurek1996cosmological}. 

In the ground state, close to the quantum critical point $g_c$, the correlation length $\xi \sim |g-g_c|^{-\nu}, \nu > 0$ and the corresponding relaxation time scales as $ \xi^z \sim |g-g_c|^{-\nu z} \sim \Delta^{-1}$ where $\Delta$ is the energy gap between the first excited state and the ground state. When the parameter $g$ is slowly varied in time, due to diverging relaxation time near $g_c$, the system can not stay close to the ground state beyond a time scale $0 < {\hat t} < \tau$ where the time remaining until the end of the quench is of the same order as the relaxation time: $\tau -{\hat t} \sim {\hat \xi}^z$ where ${\hat \xi} \equiv \xi({\hat t})$. For linearly changing control parameter,  
\be
g(t)=g_i-(g_i-g_f)\frac{t}{\tau}~,~ 0 \leq t \leq \tau
\label{gdef}
\ee
the KZ time scale, $\tau-{\hat t}$ is then determined through 
\bea
\tau -{\hat t} &\sim& |r (\tau-{\hat t})+(g_f-g_c)|^{-\nu z}
\eea
where the quench rate, $r=\frac{g_i-g_f}{\tau}$. If $|g_f-g_c| \ll r (\tau-{\hat t})$, the above equation yields $\tau-{\hat t} \sim r^{-\frac{\nu z}{1+\nu z}}$, or ${\hat \xi} \sim r^{-\frac{\nu}{1+\nu z}}$. Thus at time ${\hat t}$, the excess defect density scales as ${\hat \xi}^{-1}$. Assuming that no time evolution occurs during ${\hat t} < t < \tau$, the defect density at the end of quench decays with quench time as \cite{zurek1996cosmological} 
\be
\delta {\cal D}(\tau) \sim \left( \frac{g_i-g_f}{\tau}\right)^{\frac{\nu}{1+\nu z}} \label{KZ}
\ee
In the above argument, it was assumed that $|g_f-g_c| \sim \xi_f^{-1/\nu} \ll r (\tau-{\hat t}) \sim {\hat \xi}^{-1/\nu}$, or $\tau-{\hat t} \ll \tau_f$ where $\tau_f \sim \xi_f^z$ is the relaxation time to the final quench point. Thus,  the KZ scaling law (\ref{KZ}) holds for $1 \ll \tau \ll (g_i-g_f)\xi_f^{\frac{1+\nu z}{\nu}}$. 

For larger quench times, we can appeal to an adiabatic perturbation theory \cite{polkovnikov2005universal,de2010adiabatic,polkovnikov2011colloquium}, which shows that for  finite time quenches within the gapped phase, the approach to the ground state follows a universal $\tau^{-2}$ law, while for quenches through a critical point, the  KZ scaling law  (\ref{KZ}) holds. 
This discussion thus suggests that for quenches ending at the critical point (where $\xi_f \to \infty$) or in which the critical point is crossed, the slow quench behavior is described by the KZ scaling, while for quenches within the gapped phase, there is a crossover from KZ scaling to the adiabatic scaling when the quench time $\sim (g_i-g_f)\xi_f^{\frac{1+\nu z}{\nu}}$. 
In the following sections, working with transverse field Ising model as a test case for our argument, we provide analytical and numerical evidence for the crossover. 


\section{Model and its ground state}

We consider the one-dimensional transverse field Ising model \cite{sachdev1999quantum} defined by the Hamiltonian, 
\be
H=-J  \sum_{j=1}^N \left(\sigma_j^z \sigma_{j+1}^z+g \sigma_j^x\right)    
\label{hamiltonian}
\ee
where $\sigma^z$ and $\sigma^x$ denote the Pauli matrices. Here, we assume  ferromagnetic interactions between the spins so that $J > 0$ and restrict the external magnetic field in the transverse direction to $g\geq 0$; we also assume periodic boundary conditions for the system with $N$ sites. In the thermodynamic limit $N\to \infty$ and at zero temperature, this model exhibits a phase transition in the ground state at $g_c=1$, separating a ferromagnetic phase where the longitudinal magnetization 
$\langle\sigma_j^z\rangle \neq 0$ and a paramagnetic phase where $\langle\sigma_j^z\rangle = 0$ \cite{sachdev1999quantum}. Below we summarize the known results pertinent to our discussion, and for details, we refer the reader to \cite{mbeng2024quantum}. 

The Hamiltonian (\ref{hamiltonian}) can be diagonalized by mapping it to the spinless fermionic Hamiltonian via Jordon-Wigner transformation, $\sigma_j^x = 1-2c_j^\dagger c_j, \sigma_j^z = (c_j^\dagger + c_j) \prod_{i=1}^{j-1} (1-2c_i^\dagger c_i) $ where $c_j$'s are the fermionic operators.  
In momentum space, we then obtain
\be
H =  \sum_{k>0} 2 J (g-\cos k)(c_k^\dagger c_k -c_{-k}c_{-k}^\dagger) -2 J i \sin k(c_{k}^\dagger c_{-k}^\dagger -c_{-k}c_k )
\label{hamfou}
\ee
where $c_k = \frac{1}{\sqrt{N}}\sum_j e^{-ikj}c_j$ and the momenta $k=\pm \frac{2 m \pi}{N}, m=0, 1, ...., \frac{N}{2}-1$ for periodic boundary conditions, $c_{N+1}=c_1$. Equation (\ref{hamfou}) describes  an ensemble of two level systems with  Hamitonian $H_k$ given by the summand on the RHS of the above equation. The Hamitonian $H_k$ can be diagonalized through a Bogoliubov transformation to fermionic annihilation operator, $\gamma_k=v_k c_k - u_k c_{-k}^\dagger$ in terms of which $c_k=v_k \gamma_k +u_{k} \gamma_{-k}^\dagger$, and we obtain $H = \sum_k \epsilon_k \gamma_k^\dagger \gamma_k$, where
\be
\epsilon_k= \pm 2 J \sqrt{1+g^2-2g\cos k}
\label{disrel}
\ee
In the ground state, the energy of the $k$th mode is given by the negative root of (\ref{disrel}), and the energy gap between the ground state and the first excited state vanishes at $g_c$ as $\Delta \sim  |g-g_c|^{z\nu}$ where, $z=\nu=1$ \cite{sachdev1999quantum}. As the ground state $| \emptyset\rangle$ of the Hamiltonian $H$ must satisfy the condition $\gamma_k|\emptyset\rangle = 0$ for all $k$, one can write 
\be
| \emptyset\rangle=\prod_k (v_{k}+u_k c_{k}^\dagger  c_{-k}^\dagger)  |0 \rangle \label{GSpsi}
\ee
where $|0\rangle$ is the vacuum state of the original fermions ($c_k|0\rangle=0$),  and $u_k$ and $v_k$ are given by 
\bea
\left(\begin{matrix}
    u_{k} \\
    v_{k} \\
\end{matrix} \right) = \frac{1}{\sqrt{2\epsilon_k(\epsilon_k+a_k)}} \left( \begin{matrix}
    ib_k \\
   \epsilon_k+a_k \\
\end{matrix} \right) \label{inicond}
\eea
with $a_k =2 J (g-\cos k), b_k = 2 J \sin k$. \\
We are interested in two observables, namely, mean longitudinal defect density and mean transverse magnetization. The former can be obtained using Kramers-Wannier duality which maps the Hamiltonian (\ref{hamiltonian}) with parameters $(J, g)$ to $(J', g')$ where, throughout the manuscript, the superscript prime refers to quantities with $J'=J g, g'=1/g$. We then obtain the equilibrium mean defect density to be 
\bea
{\cal D}^z_{eq}(g) &=& \frac{1}{2 N} \sum_{i=1}^{N} \langle \emptyset|(1-\sigma^z_i \sigma^z_{i+1}) | \emptyset \rangle \\
&=& \frac{1}{N} \sum_{k } |u'_k|^2\label{Ldddefn} 
\stackrel{N \to \infty}{\approx} \frac{1}{2\pi} \int_{-\pi}^{\pi} dk \frac{\epsilon'_k-a'_k}{\epsilon'_k} \\
&=&
    \frac{\pi+(g-1) K\left(\frac{4 g}{(g+1)^2}\right)-(g+1) E\left(\frac{4 g}{(g+1)^2}\right) }{2 \pi }
\label{Deq}
\eea
where, $E(x)$ and $K(x)$, respectively, are the complete elliptic integrals of first and second kind. Similarly, the equilibrium mean transverse magnetization is given by 
\bea
M^x_{eq}(g)&=&  \frac{1}{N} \sum_{i=1}^N \langle \emptyset | \sigma_i^x| \emptyset \rangle \\
&=&\frac{1}{N}\sum_k (1-2 |u_k|^2) \label{magdef}
\stackrel{N \to \infty}{\approx}  \frac{1}{2\pi} \int_{-\pi}^\pi dk \frac{a_k}{\epsilon_k} \\
&=&   \frac{(g-1) K\left(\frac{4 g}{(g+1)^2}\right)+(g+1) E\left(\frac{4 g}{(g+1)^2}\right)}{g \pi } 
    \label{magx0} 
\eea

\section{Dynamics}

\subsection{Exact solution}

We now consider the Hamiltonian (\ref{hamfou}) with arbitrary time-dependent field $g(t)$. Working in the Heisenberg picture and  writing $c_k(t) = v_k(t) \gamma_k-u_{k}(t) \gamma^{\dagger}_{-k}$, we obtain the time evolution equations for the coefficients $u_k(t)$ and $v_k(t)$ which are given by \cite{dziarmaga2005dynamics}
\bea
i \frac{du_k}{dt} &=& -2 J (g(t) -\cos k) u_k + 2 i J \sin k \; v_k \label{ukvk2}\\
i\frac{dv_k}{dt} &=& -2 i J \sin k\; u_k + 2 J (g(t) -\cos k) v_k
\label{ukvk}
\eea
Then, as in (\ref{Ldddefn}), the time-dependent mean longitudinal defect density can be written in terms of $u_k'(t)$, and is given by 
\bea
\mathcal{D}^z(t)
&=& \frac{1}{2 \pi} \int_{-\pi}^{\pi} dk \left| \cos \left(\frac{k}{2} \right) {\tilde u}_k(t) - i \sin \left(\frac{k}{2}\right) {\tilde v}_k(t) \right|^2
\label{def_def}
\eea
where, 
\bea
{\tilde u}_k &=& \cos \left(\frac{k}{2}\right)  {u}'_k +i \sin  \left(\frac{k}{2}\right) {v}'_k \label{uutilde} \\
{\tilde v}_k &=& i \sin \left(\frac{k}{2}\right){u}'_k + \cos \left(\frac{k}{2}\right) {v}'_k  \label{vvtilde}
\eea
so that 
\bea
i \frac{d{\tilde u_k}}{dt} &=&  2 J (g(t) - \cos k) {\tilde u_k} + 2 i J \sin k \;{\tilde v_k} \label{uktdef}\\
i\frac{d{\tilde v_k}}{dt} &=& - 2 i J \sin k \;{\tilde u_k} - 2 J (g(t)- \cos k) {\tilde v_k}
\label{vktdef}
\eea

Similarly, analogous to (\ref{magdef}), in the thermodynamic limit, the time-dependent mean transverse magnetization is given by
\be
M^x(t)=1- \frac{1}{\pi} \int_{-\pi}^\pi dk \; |u_k(t)|^2 
\label{magdeft}
\ee
which can be obtained from the solution of (\ref{ukvk2}) and (\ref{ukvk}). 
On comparing (\ref{ukvk2}) and (\ref{ukvk}) for $u_k, v_k$ with  (\ref{uktdef}) and (\ref{vktdef})  for 
${\tilde u}_k, {\tilde v}_k$, we note that they are related via $u_k \leftrightarrow {\tilde v}_k, v_k \leftrightarrow -{\tilde u}_k$. 

For linear quenches defined by (\ref{gdef}), the exact solution of (\ref{ukvk2}) and (\ref{ukvk}) [and the corresponding equations for ${\tilde u_k}$ and ${\tilde v_k}$]  can be written in terms of parabolic cylinder functions \cite{dziarmaga2005dynamics}, and is described in Appendix~\ref{App0}. 
We assume that the system is initially prepared in the ground state with field $g_i$ so that these coupled equations are  subject to initial conditions, 
\be
{u_k}(0)={u}_{k, eq}(g_i), {v_k}(0)={v}_{k, eq}(g_i) \label{ictf}
\ee
where, ${u}_{k, eq}, {v}_{k, eq}$ are given by  (\ref{inicond}). But for $g_i \gg 1$, as is assumed here, we may write
\bea
    \begin{aligned}
        u_{k,eq} &\approx \frac{i \sin k}{2 g_i}, & {u'}_{k,eq} &\approx i  \cos \frac{k}{2}, & {\tilde u}_{k,eq} &\approx i \\
        v_{k,eq} &\approx 1, & {v'}_{k,eq} &\approx \sin \frac{k}{2}, & {\tilde v}_{k,eq} &\approx {\frac{\sin k}{2g_i}}
    \end{aligned}
\eea
These approximations allow one to simplify the expressions for $u_k(t), v_k(t)$ and their dual counterparts, and are given in Appendix~\ref{App1}. In the following discussion, we set $J=1$ in all the expressions and figures. 

\subsection{Adiabatic perturbation theory}
\label{APT}

For a slowly changing Hamiltonian $H(t)$, its eigenstate $|\psi(t) \rangle$ can be expanded in the basis of instantaneous eigenstates, that is, 
$ | \psi(t) \rangle= \sum_n a_n(t) e^{-i \int_0^t \epsilon_n(t') dt'}  | \phi_n(t) \rangle$ where $H(t) | \phi_n(t) \rangle=\epsilon_n(t) | \phi_n(t) \rangle$. Here, we use the approach of \cite{rigolin2008beyond} to determine the coefficients $a_n$ for the TFIM, as explained in Appendix~\ref{app_APT}. 

As already mentioned, due to (\ref{hamfou}), the Hamiltonian, $H(t)=\sum_k H_k(t)$ so that the  expectation of an operator $O$ with respect to $|\psi(t) \rangle$ can be written as 
 $\langle \psi(t) |O|\psi(t) \rangle  = \frac{1}{2 \pi} \int_{-\pi}^\pi dk \av{O_k}$ where $\av{}$ denotes the expectation with respect to the eigenfunction $|\psi_k(t) \rangle$ of the two-level system. 
 On expanding the eigenstate $|\psi_k(t) \rangle$ in a power series in the small parameter, $r=\frac{g_i-g_f}{\tau}$, and keeping terms to quadratic order,  we obtain
 \bea
 \av{O_k}-\av{O_k}_{eq} &=& r [\langle\psi_k^{(0)}|O_k|\psi_k^{(1)}\rangle+h.c.] + r^2 [\langle\psi_k^{(0)}|O_k|\psi_k^{(2)} \rangle+h.c.+\langle\psi_k^{(1)}|O_k|\psi_k^{(1)}\rangle] \nonumber \\
&-& r^2 \av{O_k}_{eq} [\langle\psi_k^{(0)}|\psi_k^{(2)} \rangle+h.c.+\langle \psi_k^{(1)}|\psi_k^{(1)}\rangle] 
\label{apt}
\eea
where $|\psi_k^{(p)}\rangle$ denotes the eigenfunction of the Hamiltonian $H_k$ in the $p$th order perturbation theory which are obtained in Appendix~\ref{app_APT}, and $\av{O_k}_{eq}$ gives the expectation of $O_k$ with respect to the ground state. 
 


\section{Mean defect density in paramagnetic phase}


\begin{figure}[t]
     \centering
    \begin{subfigure}{0.49\textwidth}
         \centering
         \includegraphics[width=1.0\textwidth]{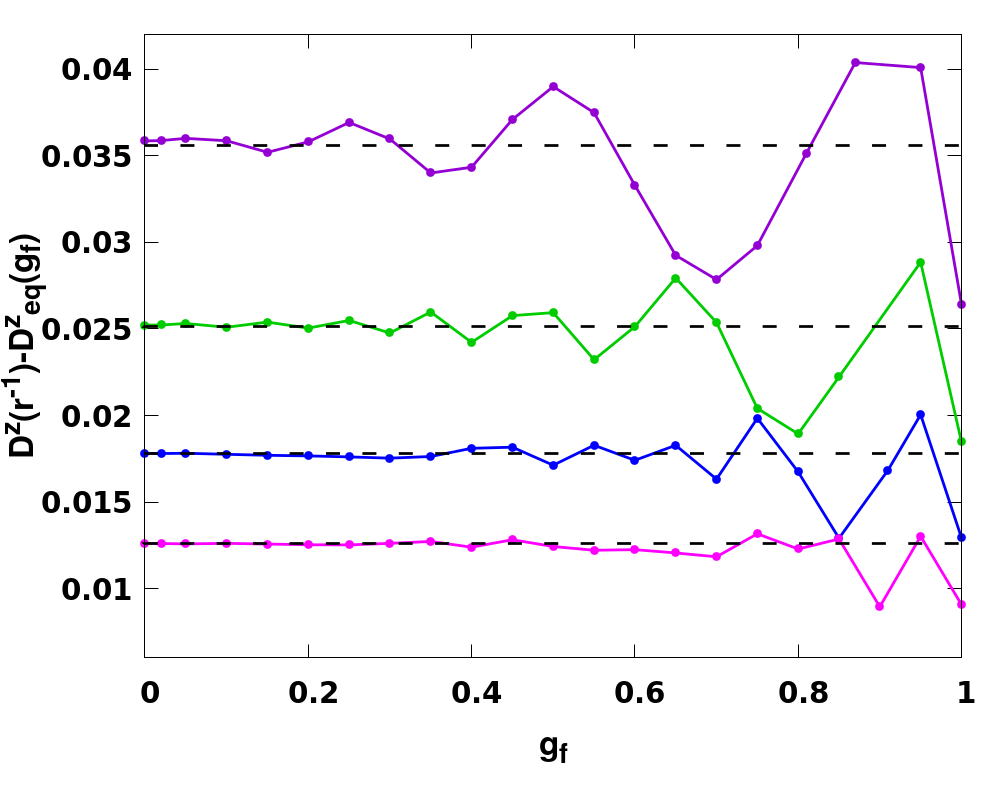}
         \caption{}
         \label{fig_fulla}
     \end{subfigure}
     \begin{subfigure}{0.49\textwidth}
         \centering
         \includegraphics[width=1.0\textwidth]{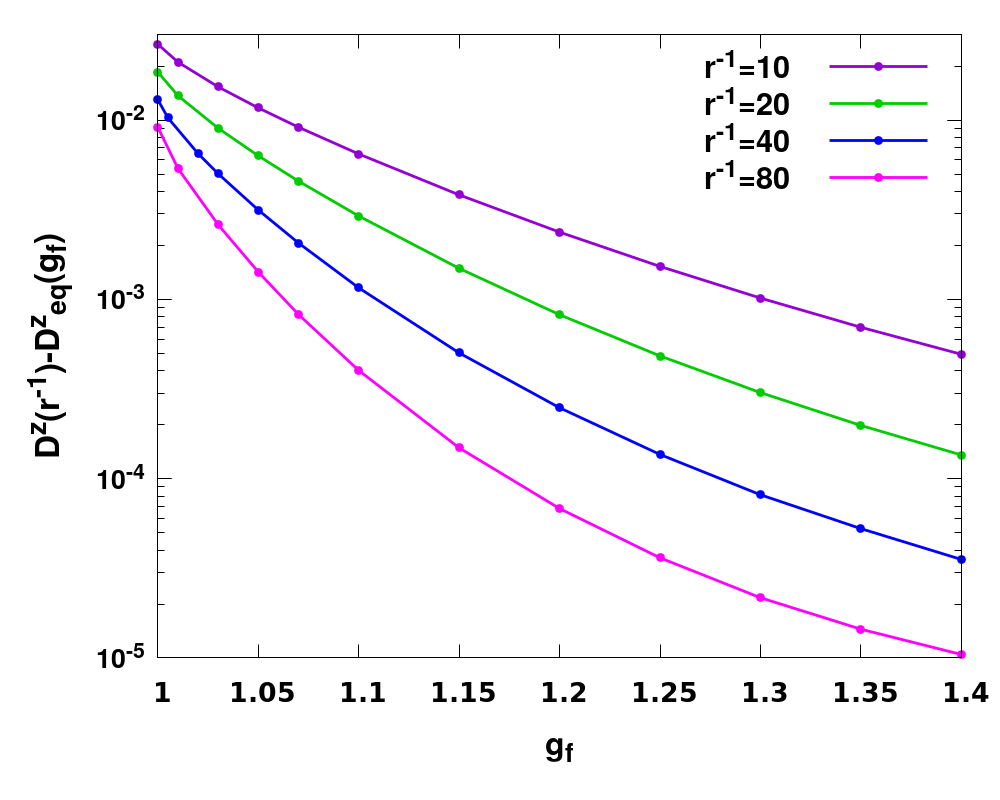}
         \caption{}
               \label{fig_fullb}
     \end{subfigure}
     \caption{Variation of excess mean longitudinal defect density at the end of quench when the system is quenched from $g_i=10^4$ to (a) $0 \leq g_f \leq 1$ and (b) $g_f \geq 1$ for $\frac{1}{r}=\frac{\tau}{g_i-g_f}=10, 20, 40, 80$ (top to bottom) in both figures. The points (joined by solids lines for clarity) are obtained using the exact results (\ref{uztsol}) and (\ref{vztsol})  for  ${\tilde u_k}$ and ${\tilde v_k}$, respectively, in (\ref{def_def}) and integrating it  numerically, and using (\ref{Deq}); the dashed lines in Fig.~\ref{fig_fulla} show the analytical result (\ref{dzresult}).} 
       \label{fig_full}
\end{figure}

In Fig.~\ref{fig_full}, we show how excess mean longitudinal defect density $\delta {\cal D}^z(\tau)={\cal D}^z(\tau)-{\cal D}^z_{eq}(g_f)$ varies with $g_f$, when the system initially in the ground state far from the critical point ($g_i \gg 1$) is   quenched according to the protocol (\ref{gdef}) to $g_f$ above, at, and below the critical point at $g_c=1$.  
When the system is quenched to the ferromagnetic phase, we find that $\delta {\cal D}^z$ oscillates with $g_f$ about a constant. But these oscillations dampen with increasing quench time, and the excess defect density is given by  \cite{dziarmaga2005dynamics}
\be
\delta {\cal D}^z(\tau) = \frac{1}{2 \pi} \sqrt{\frac{g_i-g_f}{2 \tau}}, \hspace{0.4in} 0 \leq g_f < 1 \label{dzresult} 
\ee
when quench time is large and $g_f$ is not close to the critical point. 
On the other hand, when the system is quenched within the paramagnetic phase, the excess mean defect density is far smaller than that for $g_f < 1$, as one may expect, and decreases with increasing $g_f$ and $\tau$. 

%

Here, we are interested in quenches within the gapped phase, and show the excess mean defect density when the quench ends in the paramagnetic phase in detail in Fig.~\ref{fig1a} as a function of quench time. We find that there are three distinct scaling regimes where $\delta {\cal D}^z$ is either a constant, or decays either as $\tau^{-1/2}$ or $\tau^{-2}$, depending on the quench rate, $r=\frac{g_i-g_f}{\tau}$ and the correlation length $\xi_f \sim (g_f-1)^{-1}$ in the ground state at $g_f$, as explained in Sec.~\ref{scar}.

\subsection{Stationary regime}

\begin{figure}[t]
     \centering
    \begin{subfigure}{0.495\textwidth}
         \centering
         \includegraphics[width=1.0\textwidth]{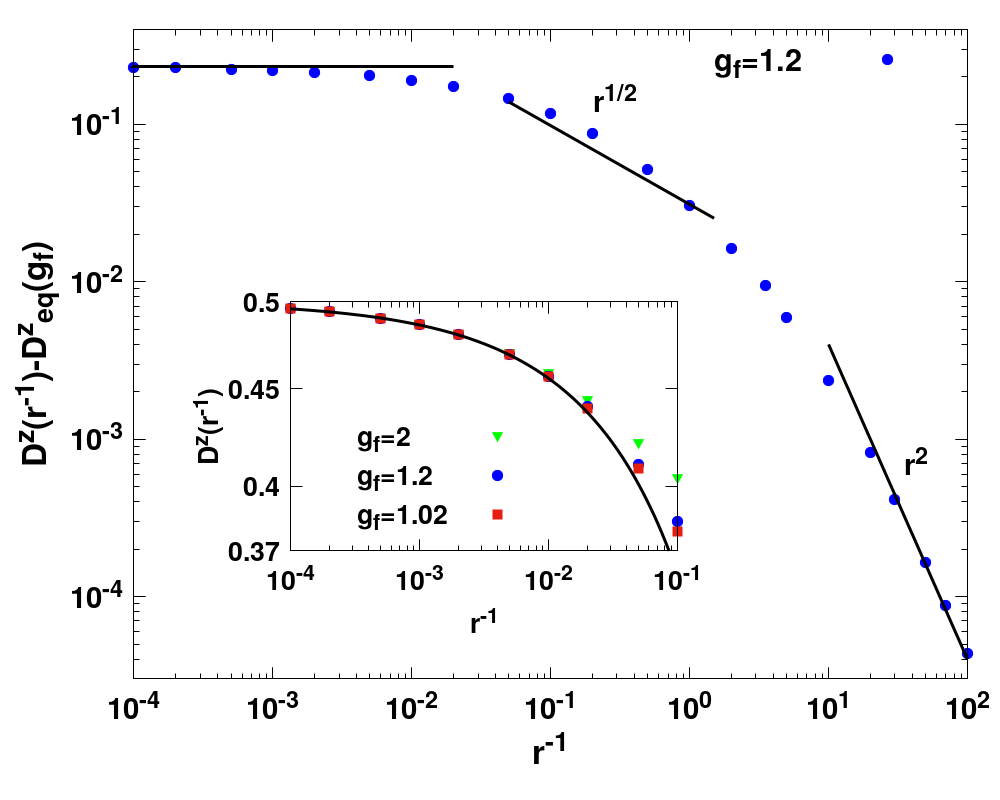}
         \caption{}
         \label{fig1a}
     \end{subfigure}
     \begin{subfigure}{0.495\textwidth}
         \centering
         \includegraphics[width=1.0\textwidth]{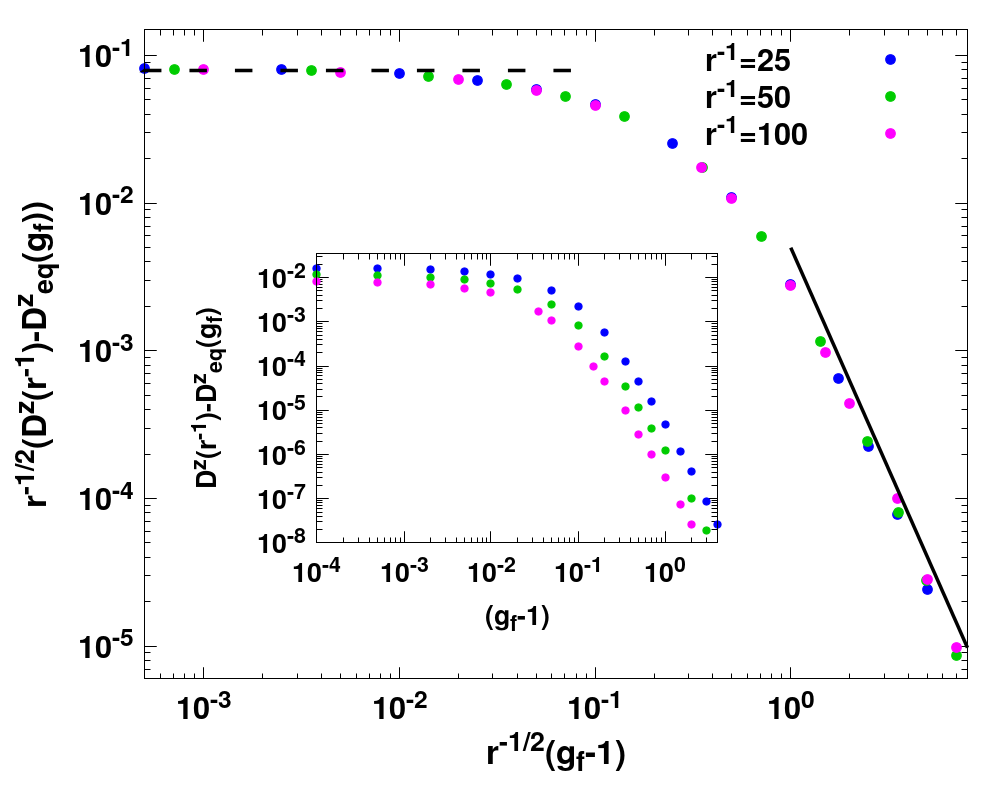}
        \caption{}
        \label{fig1b}
     \end{subfigure}
        \caption{(a) Excess mean longitudinal defect density  at the end of the quench is plotted against the inverse quench rate, $\frac{1}{r}=\frac{\tau}{g_i - g_f}$ for fixed $g_i=10^4$ and $g_f=1.2$ to show the three scaling regimes  (see text for details).  
The inset figure shows the defect density in the stationary regime, and the black solid line is the corresponding analytical expression (\ref{smallxD}). (b) The inset shows the excess defect density for $\tau \gg g_i-g_f$ in the KZ and adiabatic regime when both $g_f$ and $\tau$ are varied while the main figure shows that these data collapse onto a single scaling curve when plotted according to (\ref{scalfun}). The dashed line shows the analytical result (\ref{ddKZ}) in the KZ regime, and a numerical fit to the data in the adiabatic regime gives the slope of the solid line to be $-3$. In both the figures, the points are obtained using the exact results (\ref{uztsol}) and (\ref{vztsol})  for  ${\tilde u_k}$ and ${\tilde v_k}$, respectively, in (\ref{def_def}) and integrating it  numerically, and using (\ref{Deq}).}
        \label{fig1}
\end{figure}

For a rapid quench, that is, $\tau \ll g_i-g_f$,  the system does not have sufficient time to relax in response to the changing $g$, and stays close to its initial state so that ${\cal D}^z(\tau) \approx {\cal D}_{eq}^z(g_i)$. In Appendix~\ref{app_smallx}, using the power series expansions for the parabolic cylinder functions, we find that the mean defect density at the end of the quench is given by 
\bea
{\cal D}^z(\tau) &\approx &  \frac{1}{2} -\frac{\sqrt{\pi}}{4} \sqrt{\frac{\tau}{g_i-g_f}} \label{smallxD}
\eea
which matches well with the exact results shown in the inset of Fig.~\ref{fig1a}. For $g_f=0$, the above expression reduces to (31) of \cite{kou2023varying} where the effect of fast quenches has been investigated (up to quadratic order in $1/g_i$). 


\subsection{Kibble-Zurek scaling regime}

As argued in Sec.~\ref{scar}, for moderately slow quenches where $1 \ll \frac{\tau}{g_i-g_f} \ll \frac{1}{(g_f-1)^2}$, the excess defect density is expected to decay according to the KZ scaling law (\ref{KZ}). The excellent data collapse in the inset of Fig.~\ref{fig1b} attests that $\delta {\cal D}^z$ is of the following scaling form:
\be
\delta {\cal D}^z(\tau, g_f)= \sqrt{\frac{g_i-g_f}{\tau}} F \left( (g_f-1) \sqrt{\frac{\tau}{g_i-g_f}} \right)
\label{scalfun}
\ee
where the scaling function $F(y)$ is constant for $y \ll 1$ which means that the KZ scaling law (\ref{KZ}) holds for moderate quench rates.
As shown in Appendix~\ref{app_KZ}, the excess mean defect density at the end of the quench is given by 
\bea
\delta{\cal D}^z(\tau) &\approx&  \frac{1}{4 \pi} \sqrt{\frac{g_i-g_f}{\tau}} \label{ddKZ}
\eea
which matches well with the exact results in Fig.~\ref{fig1b}. Note that the above excess mean defect density for the quench from  deep in the paramagnetic phase to just above the critical point is $1/\sqrt{2} \approx 0.707$ times smaller than that for the quench from paramagnetic to ferromagnetic phase [see (\ref{dzresult})]. 

\subsection{Adiabatic scaling regime}
\label{ddAPT}

For slow quenches where $\frac{\tau}{g_i-g_f} \gg \frac{1}{(g_f-1)^2}$, the excess mean defect density decays as $\tau^{-2}$, or equivalently, the scaling function [see (\ref{scalfun}) above], $F(y) \sim y^{-3}$ as shown in Fig.~\ref{fig1b}.  
The calculation of the exact prefactor for the excess mean defect density seems too tedious and is not pursued here; however, in the following section, we find the exact amplitude for the mean transverse  magnetization in the adiabatic scaling regime.



\section{Mean magnetization in paramagnetic phase}

We again consider the situation when the system is prepared in the ground state at $g_i\gg 1$ and slowly quenched to $g_f > 1$ using the protocol (\ref{gdef}). The mean transverse magnetization at the end of quench is obtained using (\ref{uzsol}) and (\ref{vzsol}) in (\ref{magdeft}), and we find that the behavior of the excess mean transverse magnetization, $\delta M^x=M^x(\tau)-M_{eq}^x(g_f)$ can be classified in three scaling regimes as discussed below.
\subsection{Stationary regime}
\label{smallq}

As shown in Appendix~\ref{app_smallx}, for $\tau \ll g_i-g_f$, using a power series expansion of the parabolic cylinder functions, the transverse magnetization at the end of the quench is found to be 
\bea
M^x (\tau) 
&\approx& 1-\frac{\pi}{2} \frac{\tau}{g_i-g_f}
\label{magsmallx}
\eea
which is in agreement with the data shown in the inset of Fig.~\ref{fig_maga}. 

\subsection{Kibble-Zurek scaling regime}

\begin{figure}[t]
     \centering
    \begin{subfigure}{0.495\textwidth}
         \centering
         \includegraphics[width=1.0\textwidth]{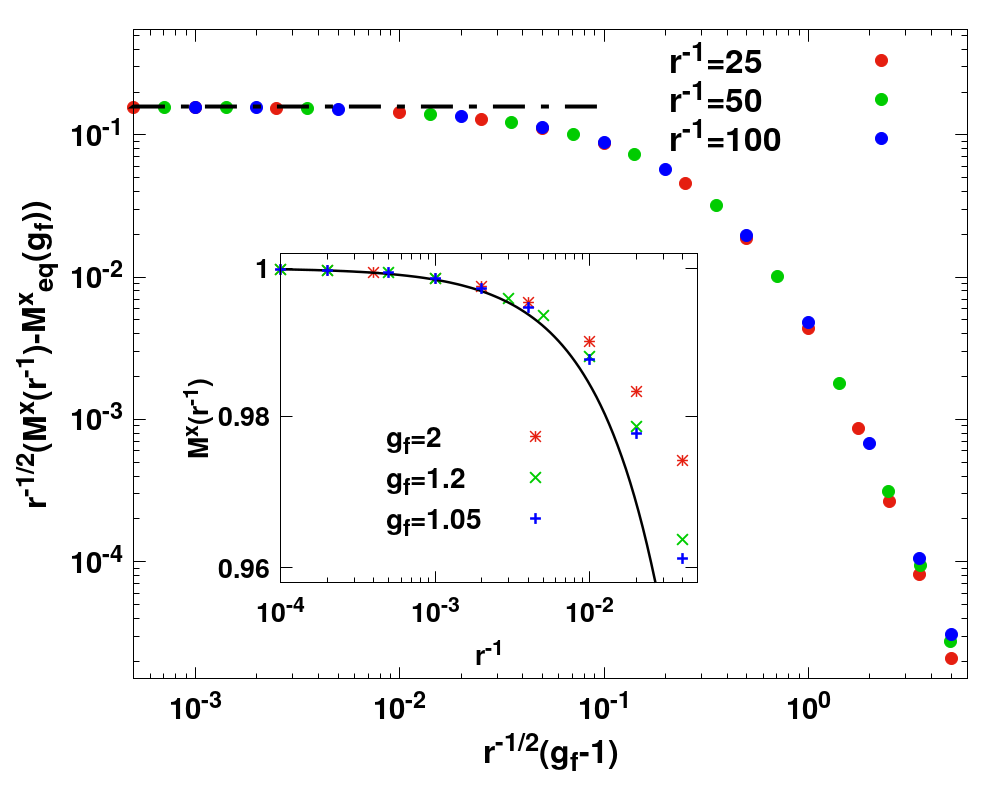}
         \caption{}
         \label{fig_maga}
     \end{subfigure}
     \begin{subfigure}{0.495\textwidth}
         \centering
         \includegraphics[width=1.0\textwidth]{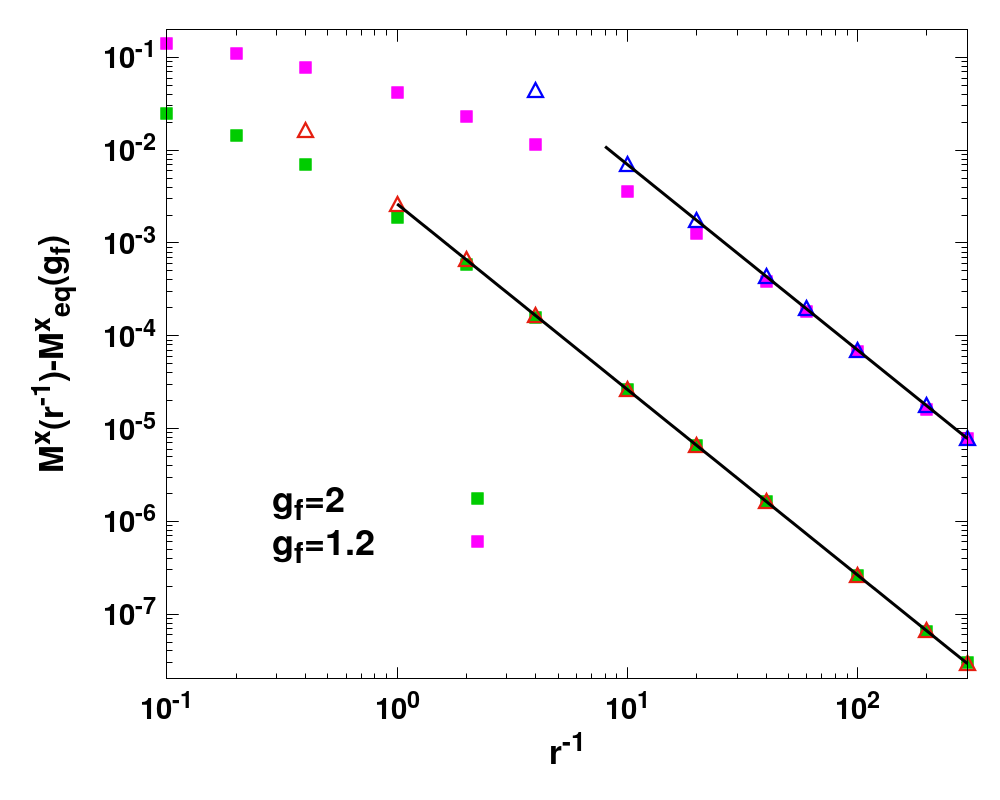}
        \caption{}
        \label{fig_magb}
     \end{subfigure}
        \caption{(a) The main figure shows the data collapse for the excess mean transverse magnetization at the end of quench for $\tau\gg g_i-g_f$ in the KZ and adiabatic regime when the system is quenched from $g_i=10^4$ to $g_f$ for various inverse quench rate, $\frac{1}{r}=\frac{\tau}{g_i-g_f}$. The dashed line shows the analytical result (\ref{magKZ}) in the KZ regime, whereas the inset figure shows the mean magnetization in the stationary regime and black solid line is the corresponding analytical expression (\ref{magsmallx}). (b) The figure shows the excess mean magnetization in the adiabatic regime for two $g_f$ values. The {open triangles} correspond to APT results and the black solid lines are the analytical result (\ref{largexmag}). In both the figures, the filled points are obtained using the exact results (\ref{uzsol}) and (\ref{vzsol}) for  $ u_k$ and $ v_k$, respectively, in (\ref{magdeft}) and integrating it numerically, and using (\ref{magdef}).}
        \label{fig_mag}
\end{figure}

For $1 \ll \frac{\tau}{g_i-g_f} \ll \frac{1}{(g_f-1)^2}$, as shown in Appendix~\ref{app_KZ}, we find that the excess transverse magnetization is given by 
 \bea
\delta M^x(\tau) \approx  \frac{1}{2 \pi} \sqrt{\frac{g_i-g_f}{\tau}}
\label{magKZ}
\eea
which displays KZ scaling. This expression is found to be in good agreement with the exact results in Fig.~\ref{fig_magb}  for intermediate quench times.


\subsection{Adiabatic scaling regime}
\label{magad}

For very slow quenches, $\frac{\tau}{g_i-g_f} \gg \frac{1}{(g_f-1)^2}\gg 1$, the system is expected to be close to the ground state at the end of the quench so that $\delta M^x$ is close to zero, and the excess magnetization decays quadratically with the quench rate. This behaviour can be explained by an adiabatic perturbation theory which is explained in 
Sec.~\ref{APT}; using the magnetization operator, $O_k=1-2c_k^{\dagger}c_k$ in (\ref{apt}), we obtained the excess transverse magnetization  numerically as the integrals over the momenta do not appear to be exactly doable.  
 
We also obtained an expression for the excess transverse magnetization at the end of the quench using uniform asymptotic expansions for the parabolic cylinder functions as described in Appendix~\ref{Bapp}), 
and find that to leading order in $1/\tau$, 
\be
\delta M^x(\tau)= A_M \left(\frac{g_i-g_f}{\tau} \right)^2
\label{largexmag}
\ee
where, the amplitude, 
\be
A_M=\frac{\left(2-3 g_f^2\right) (g_f-1)^2 K\left(\frac{4 g_f}{(g_f+1)^2}\right)+\left(3 g_f^4+7 g_f^2-2\right) E\left(\frac{4 g_f}{(g_f+1)^2}\right)}{96 \pi (g_f-1)^3 g_f^3 (g_f+1)^2 }
\ee
and $E(x)$ and $K(x)$ are the elliptic integrals of first and second kind, respectively. In Fig.~\ref{fig_magb}, we compare the exact numerical result for excess defect density with the corresponding results obtained using the adiabatic perturbative expansion (\ref{apt}) and the analytical expression (\ref{largexmag}), and find a good agreement between the three curves at sufficiently large quench times.

\section{Discussion}

In slowly quenched systems that show a second order quantum phase transition in the ground state, much work has been done to understand the residual mean defect density, kink-kink correlation function, excess energy, etc. when the system is quenched to or through the critical point \cite{zurek2005dynamics,polkovnikov2005universal,dziarmaga2005dynamics,sen2008defect,puskarov2016time,bialonczyk2018one,puebla2020kibble,roychowdhury2021dynamics,soriani2022three,zeng2023universal,kou2023varying}. Here,  we have focused on quenches within the gapped phase, and argued that with increasing quench time $\tau$, there is a crossover from the KZ scaling to the adiabatic $\tau^{-2}$ scaling when the KZ length becomes longer than the equilibrium correlation length at the end of the quench. While the latter scaling behavior has been shown using an adiabatic perturbation theory for quenches in the gapped phase \cite{polkovnikov2005universal,de2010adiabatic}, the fact that it is preceded by KZ scaling at intermediate quench rates does not appear to have been noted in previous work. 

\begin{figure}[t]
\centering
 \includegraphics[width=0.6\textwidth]{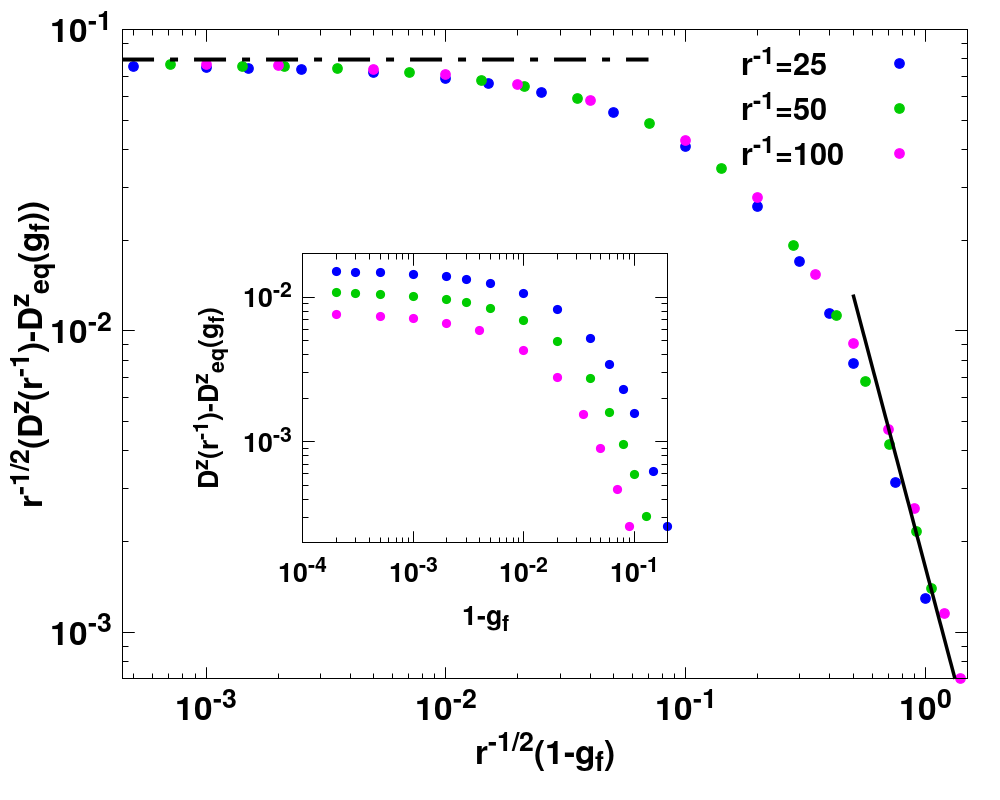}
\caption{Excess mean longitudinal defect density (\ref{def_def}) at the end of the quench for $\tau \gg g_i-g_f$ in the KZ and adiabatic regime when the system is quenched from $g_i=0$ to $0 < g_f < 1$. The inset figure shows the data when both $g_f$ and $\tau$ are varied while the main figure shows that these data collapse onto a single scaling curve according to (\ref{scalfun}). The dashed line shows the analytical result (\ref{ddKZ}) in the KZ regime, and a numerical fit to the data gives the slope of the solid line to be $-3$ in the adiabatic regime. Here, the points are obtained numerically using the exact results (\ref{uztsol}) and (\ref{vztsol})  for  ${\tilde u_k}$ and ${\tilde v_k}$, respectively, and integrating (\ref{def_def})  numerically, and using (\ref{Deq}).} 
\label{figff}
\end{figure}

In \cite{puskarov2016time}, the finite-time quench dynamics of the transverse field Ising model are studied, mainly numerically, for various quench protocols, and for a quench within the ferromagnetic phase, only $\tau^{-2}$ scaling  was noted (see their Fig.~5, left panel). 
Here, we have shown numerically and analytically that the aformentioned crossover occurs in the  paramagnetic phase of the TFIM, and Fig.~\ref{figff} shows that this result holds for slow quench within the ferromagnetic phase as well. Our analytical results in the paramagnetic phase are obtained using different kinds of expansions of the parabolic cylinder function in different parameter regimes. In particular, here we have employed the uniform asymptotic expansion for  the calculations in the adiabatic regime, the analyses of which, to our knowledge, have so far been carried out via an APT only \cite{polkovnikov2005universal}. 


The crossover behavior mentioned above is not limited to quantum systems. For the classical Glauber Ising chain \cite{jindal2024kibble} for which the correlation length remains finite at all temperatures above zero, on quenching the system from a high temperature to a low but nonzero temperature, we find that the excess mean defect density follows KZ scaling law for intermediate quench times and a nonuniversal power law for larger quench times (details will be discussed elsewhere). Thus, in general,  in an infinitely large system, for a quench ending at a finite distance from the critical point, we expect such a crossover as there are two length scales, namely, the KZ length and the correlation length in the 
ground or equilibrium state at the final quench point; see Sec.~\ref{scar} for a scaling argument. 



%
%


In this work, we focused on the role of the final value of the control parameter and assumed that the system is prepared in the ground state far from the critical point, but it would be interesting to study the effect of different initial conditions in the vicinity of the critical point. In the classical Glauber Ising chain, it has been shown that if the system starts in a nonequilibrium state, the mean defect density follows coarsening laws at small quench times but when the quench times are large enough that the system can reach an adiabatic state before the impulse regime sets in, the KZ law holds \cite{jindal2024kibble}. However, in the TFIM, our  preliminary study shows that starting from a mixed initial state in the paramagnetic phase, when the system is quenched to the ferromagnetic phase, as this model is integrable \cite{PhysRevA.2.1075,vidmar2016generalized}, the system does not approach the adiabatic state and hence KZ scaling law is not observed even at very high quench times. Perhaps, other non-integrable models may show a behavior similar to that in \cite{jindal2024kibble}, and this remains an interesting open question. \\







\noindent Acknowledgements: We thank Arnab Das and Krishanu Roychowdhury for many helpful discussions during the early stages of this work. 

\clearpage

\clearpage
\appendix
\makeatletter
\renewcommand{\@seccntformat}[1]{Appendix \csname the#1\endcsname\quad}
\makeatother
\renewcommand{\thesection}{\Alph{section}}
\numberwithin{equation}{section}

\section{Exact solution}
\label{App0}

Using the coupled equations (\ref{uktdef}) and (\ref{vktdef}), we find that $\tilde{u_k}$ and $\tilde{v_k}$ obey the following differential equations,
\bea
\frac{d^2{\tilde u_k}}{dz^2} + \left(n {+} \frac{1}{2}-\frac{z^2}{4}\right){\tilde u_k} &=&0  \label{weberu}\\
\frac{d^2{\tilde v_k}}{dz^2} + \left(n {-} \frac{1}{2}-\frac{z^2}{4}\right){\tilde v_k} &=&0  \label{weberv}
\eea
where 
\bea
z(t)&=&2\sqrt{\frac{J\tau}{g_i-g_f}}\left(g_i-(g_i-g_f)\frac{t}{\tau}-\cos k\right)e^{-i\pi/4} \\
n&=&\frac{iJ\tau}{g_i-g_f}\sin^2 k \label{ndef1}
\eea
The solution of above differential equations can be written in terms of parabolic cylinder functions $D_{n}(z)$ \cite{DLMF}, 
\bea
{\tilde u_k(z)} &=& {\tilde A}(n) D_{-n-1}(iz) + {\tilde B}(n) D_{-n-1}(-iz) \label{uztsol}\\
{\tilde v_k(z)} &=& { \kappa} {\tilde A}(n) D_{-n}(iz) {-} { \kappa} {\tilde B}(n) D_{-n}(-iz) \label{vztsol}
\eea
where,
\bea
 \kappa &=& \sqrt{\frac{g_i-g_f}{J\tau}}\frac{1}{\sin k}{e^{i\pi/4}} 
\eea
and 
 the constants ${\tilde A}(n)$ and ${\tilde B}(n)$ are found using the initial conditions to be 
\bea
{\tilde A}(n) &=& \frac{1}{D_{-n-1}(iz_i)}\Bigg(\frac{{\frac{D_{-n}(-iz_i)}{D_{-n-1}(-iz_i)}\tilde u_{k,eq}}{+}{\kappa}^{-1}{\tilde v_{k,eq}}}{\frac{D_{-n}(-iz_i)}{D_{-n-1}(-iz_i)}+\frac{D_{-n}(iz_i)}{D_{-n-1}(iz_i)}}\Bigg)  \label{Ant}\\
{\tilde B}(n) &=& \frac{1}{D_{-n-1}(-iz_i)}\Bigg(\frac{{\frac{D_{-n}(iz_i)}{D_{-n-1}(iz_i)}\tilde u_{k,eq}}{-}{ \kappa}^{-1}{\tilde v_{k,eq}}}{\frac{D_{-n}(-iz_i)}{D_{-n-1}(-iz_i)}+\frac{D_{-n}(iz_i)}{D_{-n-1}(iz_i)}}\Bigg)
\label{Bnt}
\eea
and $z_i \equiv z(0), z_f \equiv z(\tau)$. In the above equations, $\tilde u_{k,eq}$ and $\tilde v_{k,eq}$ are obtained using (\ref{uutilde}) and (\ref{vvtilde}) where, 
\bea
u'_k=\frac{i b'_k}{\sqrt{2 \epsilon'_k (\epsilon'_k+a'_k)}} , v'_k=\frac{\epsilon'_k+a'_k}{\sqrt{2 \epsilon'_k (\epsilon'_k+a'_k)}} 
\eea
since $b_k'=2 J g \sin k, a_k'=2 J (1- g \cos k) , \epsilon_k' = \epsilon_k$.

On $u_k \leftrightarrow {\tilde v}_k, v_k \leftrightarrow -{\tilde u}_k$ in 
(\ref{weberu}) and (\ref{weberv}), the corresponding equations for $u_k$ and $v_k$ are obtained with the respective solutions,
\bea
{ u_k(z)} &=& \kappa { A}(n) D_{-n}(iz) - \kappa { B}(n) D_{-n}(-iz) \label{uzsol} \\
{v_k(z)} &=& -{A}(n) D_{-n-1}(iz) - { B}(n) D_{-n-1}(-iz) \label{vzsol}
\eea
where,
\bea
{A}(n) &=& \frac{1}{D_{-n-1}(iz_i)}\Bigg(\frac{{\frac{D_{-n}(-iz_i)}{D_{-n-1}(-iz_i)} (-v_{k,eq}})+\kappa^{-1}{ u_{k,eq}}}{\frac{D_{-n}(-iz_i)}{D_{-n-1}(-iz_i)}+\frac{D_{-n}(iz_i)}{D_{-n-1}(iz_i)}}\Bigg)  \label{An}\\
{B}(n) &=& \frac{1}{D_{-n-1}(-iz_i)}\Bigg(\frac{{\frac{D_{-n}(iz_i)}{D_{-n-1}(iz_i)} (-v_{k,eq}})-\kappa^{-1}{u_{k,eq}}}{\frac{D_{-n}(-iz_i)}{D_{-n-1}(-iz_i)}+\frac{D_{-n}(iz_i)}{D_{-n-1}(iz_i)}}\Bigg)
\label{Bn}
\eea
with $u_{k,eq}$ and $v_{k,eq}$ given by (\ref{inicond}). 

%
%



\section{Asymptotic expansions of constants}
\label{App1}

As we are working with $\tau \gg 1$ and $g_i-g_f \gg 1$, we first define a scaling variable $x=\frac{\tau}{g_i-g_f}$ which is finite in these scaling limits. Using the definitions in Appendix~\ref{App0} (and setting $J=1$), we then obtain
\bea
z_i &=&2\tau\sqrt{x}\left(\frac{1}{x}+\frac{g_f-\cos k}{\tau}\right)e^{-i\pi/4}~ {\stackrel{\tau \gg 1}{\longrightarrow}} \frac{2 \tau}{\sqrt{x}} e^{-i\pi/4} \label{z0def}\\
n &=& i x \sin^2 k \label{ndef}\\
{\kappa}&=& \frac{1}{\sqrt{x} \sin k}e^{i\pi/4} \\ 
z_f &=& 2\sqrt{x}\left(g_f-\cos k\right)e^{-i\pi/4}
\label{ztau}
\eea
so that the variable $n$ is finite, while $z_i$ approaches infinity with increasing quench time. 
Thus, for  fixed index $n$ and large argument $z_i$, we can use asymptotic expansions of parabolic cylinder functions in (\ref{Ant}) and (\ref{Bnt}). 

Using the notation in Sec.~12.9 of \cite{DLMF}, we have $D_{-n}(iz) \equiv U(n-\frac{1}{2},iz)$ and $D_{-n-1}(iz) \equiv U(n+\frac{1}{2},iz)$, and from (12.9.1) and (12.9.3) of \cite{DLMF} for $D_{-n}(iz)$ and $D_{-n}(-iz)$, respectively, to leading order in $|z_i|$, we have 
\bea
D_{-n-1}(iz_i) &\sim & e^{-\frac{i\pi}{4}} e^{-\frac{in\pi }{4}-\frac{i |z_i|^2}{4}} |z_i|^{-n-1} + {\cal O}(|z_i|^{-n-3})) \\
D_{-n}(iz_i) &\sim & e^{-\frac{in\pi }{4}-\frac{i |z_i|^2}{4}} |z_i|^{-n} + {\cal O}(|z_i|^{-n-2}))\\
D_{-n-1}(-iz_i) &\sim & \frac{\sqrt{2 \pi } e^{\frac{in\pi }{4}+\frac{i |z_i|^2}{4}} |z_i|^{n}} {\Gamma (1+n)} + {\cal O}(|z_i|^{-n-1})\\
D_{-n}(-iz_i) &\sim& e^{\frac{i3 \pi  n}{4}-\frac{i |z_i|^2}{4}} |z_i|^{-n} + {\cal O}(|z_i|^{n-1})
 \eea
Hence, 
\be
\frac{D_{-n}(iz_i)}{D_{-n-1}(iz_i)} \propto \tau, 
\frac{D_{-n}(-iz_i)}{D_{-n-1}(-iz_i)} \propto \frac{1}{\tau} 
\ee
using which we can approximate (\ref{Ant}) and (\ref{Bnt}) as 
\bea
{\tilde A(n)} &\stackrel{z_i\gg 1}{\approx}& \frac{\Gamma(1+n)e^{\frac{i3n\pi}{4}-\frac{i|z_i|^2}{4}+\frac{i\pi}{2}}|z_i|^{-n}}{\sqrt{2\pi}} \\
|{\tilde A(n)}|^2 &\approx& \frac{e^{-\frac{3m\pi}{2}}|\Gamma(1+im)|^2}{2\pi} = \frac{me^{-\frac{3m\pi}{2}}}{2 \sinh (m\pi)}\\
{\tilde B(n)} &\stackrel{z_i\gg 1}{\approx}&  \frac{\Gamma(1+n)e^{-\frac{in\pi}{4}-\frac{i|z_i|^2}{4}+\frac{i\pi}{2}}|z_i|^{-n}}{\sqrt{2\pi}} = e^{m\pi} {\tilde A(n)}\\
|{\tilde B(n)}|^2 &\approx& \frac{e^{\frac{m\pi}{2}}|\Gamma(1+im)|^2}{2\pi} = \frac{me^{\frac{m\pi}{2}}}{2 \sinh (m\pi)} \label{Bn2app}
\eea
where $m=x\sin^2 k$, and $|\Gamma(1+i m)|^2 =im \Gamma(i m) \Gamma(1-im)= m\pi \text{csch}(m\pi) $ on using (5.5.3) of \cite{DLMF}. Hence, (\ref{uztsol}) and (\ref{vztsol}) can be approximated by 
\bea
{\tilde u_k}(z) &\approx& {\tilde B}(n)\left(e^{-m\pi} D_{-n-1}(iz) + D_{-n-1}(-iz)\right) \label{uktapprox} \\
{\tilde v_k}(z) &\approx& { \kappa} {\tilde B}(n) \left(e^{-m\pi} D_{-n}(iz) - D_{-n}(-iz)\right) \label{vktapprox}
\eea
Similarly, for (\ref{An}) and (\ref{Bn}), we get
\bea
A(n) &=& e^{\frac{i\pi}{2}}{\tilde A(n)} \\
B(n) &=& e^{\frac{i\pi}{2}}{\tilde B(n)} 
\eea
so that (\ref{uzsol}) and (\ref{vzsol}) can be written as
\bea
u_k(z) &\approx& \kappa B(n)\left(e^{-m\pi} D_{-n}(iz) - D_{-n}(-iz)\right) \label{ukapprox}\\
v_k(z) &\approx& -B(n)\left(e^{-m\pi} D_{-n-1}(iz) + D_{-n-1}(-iz)\right) \label{vkapprox}
\eea


\section{Adiabatic perturbation theory}
\label{app_APT}

Following \cite{rigolin2008beyond}, we expand the eigenfunction $|\psi_k(t) \rangle$ of the two-level, time-dependent Hamiltonian in the instantaneous eigenstates; due to (\ref{GSpsi}), these are given by
\bea
| \phi_{0,k}(t) \rangle &=& [\cos \left(\frac{\theta_k}{2} \right)+i \sin \left(\frac{\theta_k}{2} \right) c_k^\dagger c_{-k}^\dagger] |0 \rangle\\ 
| \phi_{1,k}(t) \rangle &=& \gamma_k^\dagger \gamma_{-k}^\dagger  | \phi_{0,k}(t) \rangle=[i \sin \left(\frac{\theta_k}{2} \right)+\cos \left(\frac{\theta_k}{2} \right) c_k^\dagger c_{-k}^\dagger] |0 \rangle
\eea
with $\tan \theta_k=\frac{\sin k}{g(t) -\cos k}$. Working with $g$ (instead of time $t$) for convenience and dropping the subscript $k$ for the momenta for brevity, we can write 
\bea
|\psi(g) \rangle &=& |\psi^{(0)}(g) \rangle+r |\psi^{(1)}(g) \rangle+r^2 |\psi^{(2)}(g) \rangle+... \\
 |\psi^{(p)}(g) \rangle &=& e^{-\frac{i}{r} \omega_0(g)} \alpha_0^{(p)} |\phi_0 \rangle+e^{-\frac{i}{r} \omega_1(g)} \alpha_1^{(p)} |\phi_1 \rangle \label{APT2} \\
 \alpha_0^{(p)}(g) &=&  \alpha_{00}^{(p)}+e^{\frac{i}{r} (\omega_0-\omega_1)} \alpha_{01}^{(p)} \\
 \alpha_1^{(p)}(g) &=& e^{\frac{i}{r} (\omega_1-\omega_0)} \alpha_{10}^{(p)} + \alpha_{11}^{(p)} 
\eea
where $g(t)$ is given by (\ref{gdef}), $r=\frac{g_i-g_f}{\tau}$, $E_1(g)=-E_0(g) = \epsilon_k(g)$, $\omega_1(g)=\int_{g_i}^g dg' E_1(g')=I(g)-I(g_i)=-\omega_0(g)$ and,  due to (2.262) of \cite{gradshteyn2007}, 
\be
I(g)=\frac{g-\cos k}{2} \sqrt{1+g^2-2 g \cos k}+\frac{\sin^2 k}{2} \sinh^{-1} \left(\frac{g-\cos k}{\sin k} \right) 
\label{int}
\ee 
In the more general setting, Berry phases should be included but for the problem at hand, we have verified that these are indeed zero. Using the above ansatz in Schr{\"o}dinger's equation, and comparing terms at same order of $r$, we  obtain \cite{rigolin2008beyond}
\bea
\partial_g \alpha^{(p)}_{lm}+\sum_{n=0,1} \langle \phi_l|\partial_g|\phi_n \rangle \alpha^{(p)}_{nm} + i(E_l-E_m)\alpha^{(p+1)}_{lm} &=&0 ~,~l, m=0,1 
\eea
where $\partial_g \equiv \frac{\partial}{\partial g}$. 

As initially the system is in the ground state, $\alpha_0^{(p)}(g)=\delta_{p,0}$ and $\alpha_1^{(0)}(g)=0$. Using these in the above recursion equation, we obtain
\bea
\alpha_{01}^{(1)}(g) &=& 0 \\
\alpha_{00}^{(1)}(g) &=& \int_{g_i}^g dg' \frac{2i\sin^2 k}{\epsilon_k^5(g')}\\
\alpha_{10}^{(1)}(g) &=& \frac{\sin k}{\epsilon_k^3(g)} \\
\alpha_{11}^{(1)}(g) &=& -\frac{\sin k}{\epsilon_k^3(g_i)} \\
\alpha_{01}^{(2)}(g)&=& \frac{\sin^2 k}{\epsilon_k^3(g) \epsilon_k^3(g_i)} \\
\alpha_{10}^{(2)}(g)&=& -\frac{6i\sin k (g-\cos k)}{\epsilon_k^6(g)} + \frac{2i\sin^3 k}{\epsilon_k^3(g)} \int_{g_i}^g  \frac{dg'}{\epsilon_k^5(g')} \\
\alpha_{11}^{(2)}(g)&=& \frac{6i\sin k (g_i-\cos k)}{\epsilon_k^6(g_i)} - \frac{2i\sin^3 k}{\epsilon_k^3(g_i)} \int_{g_i}^g  \frac{dg'}{\epsilon_k^5(g')} \\
\alpha_{00}^{(2)}(g)&=& -\frac{\sin^2 k}{\epsilon_k^6(g_i)} + \int_{g_i}^g dg' \left[ \frac{12\sin^2 k (g'-\cos k)}{\epsilon_k^8(g')} - \frac{4\sin^4 k}{\epsilon_k^5(g')}\int_{g_i}^{g'} \frac{dg''}{\epsilon_k^5(g'')}\right]
\eea
Using these coefficients in (\ref{APT2}), we obtain $|\psi^{(p)}_k(g) \rangle$ for $p=1, 2$ which then allows us to find the  expectation of an operator, as defined in (\ref{apt}) with $\av{O_k}_{eq}=\langle \phi_{0,k} (\tau)|O_k| \phi_{0,k}(\tau) \rangle$. 

\section{Power series expansions in stationary regime}
\label{app_smallx}


For $x=\frac{\tau}{g_i-g_f} \ll 1$, the index $n$ and argument $z_f$ of the parabolic cylinder function that are defined, respectively,  in (\ref{ndef}) and  (\ref{ztau}) are small. Then, 
using the power series expansions of parabolic cylinder functions given by (12.4.1) of \cite{DLMF}, we can write
\bea
D_{-n-1}(\pm iz_f) &\approx& a_0 \mp a_1 \sqrt{x} \\
D_{-n}(\pm iz_f) &\approx& 1+ ib_1 x 
\eea
where $ a_0=\sqrt{\frac{\pi }{2}}, a_1=2  (g_f-\cos k)e^{i\frac{\pi}{4}}, b_1=\frac{\gamma +\ln 2}{2} \sin ^2 k-(g_f-\cos k)^2$
and $\gamma = 0.577216$ is the Euler constant. Using these in (\ref{uktapprox}) and (\ref{vktapprox}) at the end of quench, we obtain
\bea
{\tilde u_k(z_f)} &\approx& e^{-\frac{i\tau^2}{x}+\frac{i\pi}{2}} + \mathcal{O}(x)\\
{\tilde v_k(z_f)} &\approx& e^{-\frac{i\tau^2}{x}-\frac{i\pi}{4}}\frac{\sqrt{\pi x}\sin k}{\sqrt{2}} + \mathcal{O}(x)
\eea
Then using these results in (\ref{def_def}), the mean defect density at the end of the quench can be calculated  and yields (\ref{smallxD}) in the main text.
In an analogous fashion, from (\ref{ukapprox}) and (\ref{vkapprox}) at the end of quench, we have 
\bea
u_k(z_f) &\approx & e^{-\frac{i\tau^2}{x} +\frac{i\pi}{4}}\frac{\sqrt{\pi x}\sin k}{\sqrt{2}}+ \mathcal{O}(x)\\
v_k(z_f) &\approx & e^{-\frac{i\tau^2}{x}}+ \mathcal{O}(x)
\eea
on using which in (\ref{magdeft}) yields the expression (\ref{magsmallx}) for the transverse magnetization in the rapid quench regime in the main text.

\section{Power series expansions in KZ regime}
\label{app_KZ}



In the KZ regime where $1\ll x \ll \frac{1}{(g_f-1)^2} $, as only small-$k$ modes are excited during the evolution near the critical point, the parameter $n=i x \sin^2 k \approx  i x k^2$ which is finite for large $x$ and small $k$, while $z_f=2 \sqrt{x} (g_f -\cos k) \approx  2 \sqrt{x} (g_f -1+\frac{k^2}{2}) \to 0$. Note that due to small $z_f$, here we require power  series expansions of the parabolic cylinder functions, unlike in the quenches from paramagnetic to ferromagnetic phase where, as $x \to \infty$ (due to the absence of adiabatic regime), $z_f \to -\infty$ and one requires asymptotic expansions of the parabolic cylinder functions \cite{dziarmaga2005dynamics}. 

From (\ref{Ldddefn}) and (\ref{def_def}), the excess defect density at the end of the quench is given by 
\bea
\delta \mathcal{D}^z(\tau) \stackrel{ k \to 0}{\approx} \frac{1}{2\pi} \int_{-\pi}^{\pi} dk\;\; (|{\tilde u_k}|^2-|{u'_{k,eq}}|^2) \label{appdd}
\eea
Using the power series expansions of parabolic cylinder functions given by (12.4.1) of \cite{DLMF} for small $z_f$ but finite $n$, we obtain
\bea
{\tilde u_k}(z_f) &\approx& {\tilde B}(n) \frac{\sqrt{\pi } 2^{-\frac{1}{2}-\frac{i m}{2}}  \left(1+e^{-m\pi}\right)}{\Gamma \left(1+\frac{i m}{2}\right)}\\
{\tilde v_k}(z_f) &\approx& { \kappa} {\tilde B}(n) \frac{\sqrt{\pi } 2^{-\frac{im}{2}} \left(1 -e^{-m\pi}\right)}{\Gamma \left(\frac{1}{2}+\frac{i m}{2}\right)} 
\eea 
For $k \to 0$ and $g_f \to 1+$, as $|{u'_{k,eq}}|^2 \approx \frac{1}{2}$, we finally get 
 \bea
\delta \mathcal{D}^z(\tau) \approx
 \frac{1}{2\pi}\int_{-\pi}^{\pi} dk\;\; \frac{e^{-\pi x k^2}}{2} \stackrel{x \gg 1, k \to 0}{\approx} \frac{1}{4 \pi \sqrt{x}}
 \eea
which is the expression (\ref{ddKZ}) for the defect density at the end of the quench in the main text.

In the same way, from (\ref{magdef}) and (\ref{magdeft}), for the excess magnetization, we can write 
\be
\delta M^x(\tau) = \frac{1}{\pi} \int_{-\pi}^{\pi}dk \;\; (|{u_{k,eq}}|^2-|{u_{k}}|^2)
\ee
Using $|{u_{k,eq}}|^2 \stackrel{k \to 0}{\approx} \frac{1}{2}$, and 
\bea
u_k(z_f) &\approx& \kappa B(n) \frac{\sqrt{\pi } 2^{-\frac{im}{2}} \left(1 -e^{-m\pi}\right)}{\Gamma \left(\frac{1}{2}+\frac{i m}{2}\right)}\\ 
v_k(z_f) &\approx& -B(n) \frac{\sqrt{\pi } 2^{-\frac{1}{2}-\frac{i m}{2}}  \left(1+e^{-m\pi}\right)}{\Gamma \left(1+\frac{i m}{2}\right)}
\eea
we get 
\bea
\delta M^x(\tau) &\approx& \frac{1}{2 \pi}\int_{-\pi}^{\pi}dk e^{-\pi x k^2} \stackrel{x \gg 1, k \to 0}{\approx} \frac{1}{2 \pi \sqrt{x}}
\eea 
and hence (\ref{magKZ}) in the main text. 

\section{Uniform asymptotic expansions in adiabatic regime}
\label{Bapp}

In the adiabatic regime, $x \gg \frac{1}{(g_f-1)^2}\gg 1$,  as {\it both} index $n$ and the argument $z$ of parabolic cylinder functions in (\ref{ndef}) and (\ref{ztau}), respectively, are large for large $x$, we employ the uniform asymptotic expansions of these functions as described in Sec. 12.10(v) of \cite{DLMF} . 



Using the notation in Sec.~12.1 of \cite{DLMF}, we have $ D_{-n}(\pm iz_f) \equiv U(n-\frac{1}{2},\pm iz_f)$, and 
from Sec.~12.10(v) of \cite{DLMF},
\bea
 U\left(\frac{w^2}{2},\sqrt{2} wy\right) &\stackrel{w \gg 1}{\sim}& \frac{g(w)e^{-w^2\xi(y)}}{(y^2+1)^{\frac{1}{4}}} \sum_{s=0}^{\infty} \frac{\Bar{u}_s(y)}{(y^2+1)^{\frac{3s}{2}}} \frac{1}{w^{2s}} \label{Uexpn}
\eea
where,
\bea
g(w)  &\stackrel{w \gg 1}{\sim}& \frac{1}{\sqrt{2 w}}2^{\frac{w^2}{4}+\frac{1}{4}} e^{\frac{w^2}{4}} w^{-\frac{w^2}{2}}\left(1+ \frac{1}{24w^2} +\frac{1}{576 w^4} + \cdot\cdot\cdot \right) \\
\xi(y) & = & \frac{1}{2} y \sqrt{1+y^2} + \frac{1}{2} \ln{\left(y+ \sqrt{1+y^2}\right)}
\eea
From Sec.~12.10(ii) of \cite{DLMF}, the first few coefficients in the sum in (\ref{Uexpn}) are given by 
\bea
\Bar{u}_0(y) &=& 1 \\
\Bar{u}_1(y) &=& -\frac{y(y^2+6)}{24} \\
\Bar{u}_2(y) &=& -\frac{145-249y^2-9y^4}{1152}
\eea
On comparing the arguments of the functions $ U\left(\frac{w^2}{2},\sqrt{2} wy\right) \equiv U(n-\frac{1}{2},\pm iz_f)$ for $D_{-n}(\pm iz_f)$, we have
\be
w^2=2 n-1 =2 i m-1 
\ee
and 
\bea
y=\frac{i z_f}{\sqrt{2} w}= \frac{i z_f}{\sqrt{2(2 i m-1)}} 
\eea
where $m=-in = x\sin^2 k$.

For transverse magnetization at the end of the quench, from (\ref{magdeft}), we need to calculate $|u_k(z_f)|^2$. We first note that 
$D_{-n}(\pm iz_f) \stackrel{z_f, n \gg 1}{\sim} e^{\frac{m\pi}{4}}$
because 
$\left(\frac{2 e}{w^2} \right)^{w^2/4} e^{-w^2 \xi(y)} \sim (w^2)^{-i m/2}\sim e^{-(i \pi/2) (i m/2)}$
on ignoring the factors that are either a phase or algebraic in $x$. As a result, the first term on the RHS of  (\ref{ukapprox}) for $u_k$ can be neglected,  and we obtain
\bea
|u_k(z_f)|^2 \stackrel{x\gg 1}{\approx} |\kappa B(n) D_{-n}(-iz_f)|^2 \approx |D_{-n}(-iz_f)|^2
\eea
since $|\kappa B(n)|^2 = \frac{1}{x \sin^2 k} \frac{me^{\frac{m\pi}{2}}}{2\sinh(m\pi)} \stackrel{x \gg1}{\approx} e^{-\frac{m\pi}{2}}$.   Thus, we need to find $|D_{-n}(-iz_f)|^2 \equiv |U\left(\frac{w^2}{2},-\sqrt{2} wy\right)|^2 $ for large $x$.


Considering each factor on the RHS of (\ref{Uexpn}), we obtain: 
\bea
&&|g(w)|^2 \approx \frac{e^{\frac{m\pi}{2}}}{2}\left(1+\frac{g_1}{x^2}\right) ~,~g_1=-\frac{145 \csc ^4 k}{2304} \\
&&\left|\frac{1}{(y^2+1)^{\frac{1}{4}}}\right|^2  \approx \frac{b_k}{\epsilon_k} \left(1+ \frac{y_1}{x^2} \right)
,~y_1= \frac{1}{ b_k^4}-\frac{1}{\epsilon_k^4} \\
&&|e^{-w^2\xi}|^2 \approx \frac{\epsilon_k-a_k}{b_k} \left(1+\frac{\xi_1}{x^2} \right)~,~\xi_1= \frac{2a_k}{3\epsilon_k^3 b_k^4}\left(3+ 2 g_f^2-\cos k (4 g_f+\cos k)\right) \\
&&\left|1-\frac{y(y^2+6)}{24w^2(y^2+1)^{\frac{3}{2}}}\right|^2 \approx \left(1+ \frac{y_2}{x^2} \right)
\eea
where 
\bea
y_2 &=& \frac{1}{72\epsilon_k^6b_k^2}\Bigg[159-1212g_f^2-72g_f^4+12g_f \cos k(101+24g_f) +4\cos 2k(195g_f^2-154) \nonumber \\ &-& 924g_f\cos 3k+385\cos 4k + \frac{8a_k^2}{b_k^2}\left(\frac{\epsilon_k^2}{4}+5\sin^2 k\right)^2 - \frac{96a_k\epsilon_k}{b_k^2} \nonumber \\ &\times& \left(18+5 g_f^2+2 g_f^4+\cos k \Big(\cos k \left(7g_f^2+2g_f \cos k+15 \cos ^2 k-31\right)-2g_f (4g_f^2+5)\Big)\right)\Bigg]
\eea
which finally gives, 
\bea
|D_{-n}(-iz_f)|^2 &\approx& e^{\frac{m\pi}{2}}\left(\frac{\epsilon_k-a_k}{2\epsilon_k} - \frac{5a_k b_k^2}{4\epsilon_k^7}\right)
\label{Dnapp1}
\eea 
Thus, we obtain 
\bea
|u_k(z_f)|^2 - |u_{k,eq}|^2 &\approx& - \frac{5a_k b_k^2}{4\epsilon_k^7}
\eea
where $a_k = 2(g_f - \cos k)$, $b_k = 2 \sin k$ and $\epsilon_k = 2\sqrt{1+g_f^2-2g_f\cos k}$, and $u_{k,eq}$ is given by (\ref{inicond}). 




\clearpage
\bibliography{ref}

\end{document}